\documentclass[a4paper,10pt]{article}
\usepackage{mathrsfs}
\usepackage{graphicx}
\usepackage{latexsym}
\usepackage{amsmath}
\usepackage{amssymb}
\usepackage{textcomp}
\usepackage{amsbsy}
\title{\large \bf 126 GeV Higgs and ATLAS bound on the lightest graviton mass in Randall-Sundrum model}
\author{Ashmita Das\footnote{E-mail address: tpad@iacs.res.in}  and
Soumitra SenGupta\footnote{E-mail address: tpssg@iacs.res.in}\\
Department of Theoretical Physics,\\
Indian Association for the Cultivation of Science,\\
2A $\&$ 2B Raja S.C. Mullick Road,\\
Kolkata - 700 032, India.\\[20mm]}
\date{}
\begin{document}
\maketitle

\begin{abstract}
In the search for extra dimension through dilepton events 
in 7-TeV proton-proton collision, the ATLAS detector at LHC
 has set stringent lower bound on the mass of 
the Randall-Sundrum (RS) lightest graviton  Kaluza-Klein ( KK ) mode.  
Considering that the Randall-Sundrum model undertakes to resolve the
 well-known gauge hierarchy/fine tuning problem to restrict the Higgs mass 
within the estimated $\sim 126$ GeV against large radiative 
correction upto the cut-off of the model, we explore the 
allowed parameter space within which the RS model can
be trusted. We show that the consistency of the model 
with ATLAS results constrains the cut-off of the theory 
which is  atleast two order lower than the Planck/Quantum gravity scale 
implying the possible existence of a new Physics at this lower scale.



\end{abstract}
\section*{Introduction}
One of the main goal of the experiments in the Large Hadron Collider(LHC),
is the search for physics beyond standard model(BSM).
The motivations for new physics beyond standard model stems from the
large hierarchy of mass scales between the Planck and 
the TeV scales which results into the well-known fine tuning problem
in connection with the mass of the Higgs boson, the only scalar particle in the standard model. 
It has been shown that due to large radiative corrections the  Higgs 
mass can not be confined within TeV scale ( the current estimated value $\sim 126$ GeV ), unless some unnatural tuning
is done order by order in the perturbation theory.
Among several proposals to address this problem,
the models with extra spatial dimensions draw lot of attentions.
In particular the warped geometry
model proposed by Randall and Sundrum \cite{RS} has drawn special attention
for the following reasons : (1) It resolves the hierarchy problem 
without introducing any other intermediate scale in the theory,
(2) The modulus of the extra dimensional model can be stabilized \cite{GW1}, 
and (3) A warped solution, though not exactly same as RS model, can be found
from string theory which as a fundamental 
theory predicts inevitable existence of extra dimensions \cite{Green}.
Due to these unique features, experiments in LHC are designed
to explore the signature of these warped
extra dimensions through the dileptonic decay of   
Kaluza-Klein graviton KK modes present in these models.\\
In this work we propose to study some theoretical 
constraints of RS model and look for their consistency with the result 
obtained in LHC so far.\\
We begin with a brief description of RS model.
\section*{RS Model (brief description)}
Randall-Sundrum scenario \cite{RS}
which is defined on a 5-dimensional anti de-Sitter space-time with one spatial
direction orbifolded on $S^1/Z_2$  has the following features : 
\begin{itemize}
\item Two flat 3-branes namely hidden/Planck brane and 
visible/standard model brane are located at the two orbifold fixed points. 
The brane tension of the standard model/visible brane is negative. 
\item 5-dimensional Planck scale is nearly equal to 4-dimensional 
Planck scale.
\item Without introducing any extra scale, other than the Planck scale,in the theory 
one can choose the brane separation modulus $r_c$ to have a 
value $\sim M_{Pl}^{-1}$ such that the desired warping can be obtained between the
two branes from Planck scale to TeV scale. 
\item The modulus can be stabilized to the above chosen value 
by introducing scalar in the bulk \cite{GW1} without any further fine tuning.
\end{itemize}
We first briefly outline the RS model below:\\
The RS model is charecterized by the non-factorisable 
background metric,
\begin{equation}
ds^2 = e^{- 2 \sigma(\phi)} \eta_{\mu\nu} dx^{\mu} dx^{\nu} +r_c^2 d\phi^2 \label{eq1}
\end{equation}
with $\eta_{\mu\nu}=(-,+,+,+)$ and $\sigma=kr_c|\phi|$. 
$r_c$ is the compactification radius for the extra dimension
and $k$ is of the order of 4-dimensional Planck
scale $M_{Pl}$ and relates the 5D Planck scale $M$ to the cosmological constant
$\Lambda$. The extra dimensional coordinate is denoted by  $\phi$ and  ranges from
$-\pi$ to $+\pi$ following a $S^1/Z_2$ orbifolding. Two 3-branes are 
located at the orbifold fixed points $\phi=(0,\pi)$. The standard model 
fields are residing on the visible brane and only gravity can propagate 
in the bulk. Solving five dimensional Einstein's equation
and using orbifolded boundary condition,
the warped solution for the metric turns out to be,
$\sigma(\phi)=kr_c|\phi|$,
Where $k=\sqrt{\frac{-\Lambda}{24M^3}}$.
The visible and Planck brane tensions are,
$V_{hid}=-V_{vis}=24M^3k^2$.
All the dimensional parameters described above are
related to the reduced 4-dimensional Planck scale $\overline{M}_{Pl}$ as,
\begin{equation}
 \overline{M}_{Pl}^2=\frac{M^3}{k}(1-e^{-2kr_c\pi})\label{rplanckmass}
\end{equation}
For $kr_c \approx 12$, the exponential factor present
in the background metric, which is often called warp factor, produces 
a huge suppression on the Planck scale mass parameters and reduced those 
parameters to TeV scale on the visible brane. 
Thus a scalar mass say mass of Higgs is given as, 
\begin{equation}
 m_H=m_{0}e^{-kr_c\pi}\label{physmass}
\end{equation}
Here, $m_H$ is Higgs mass parameter 
on the visible brane and $m_0$ is the the cut off scale of the theory, above
which new physics beyond standard model is expected to appear. A natural choice for this
would be Planck or quantum gravity scale beyond which standard model will not be valid. 
\section*{ Theoretical restriction on $\epsilon=k/\overline{M}_{Pl}$ }
\begin{figure}[!h] 
\begin{center}
\centering
\includegraphics[width=3.0in,height=1.70in]{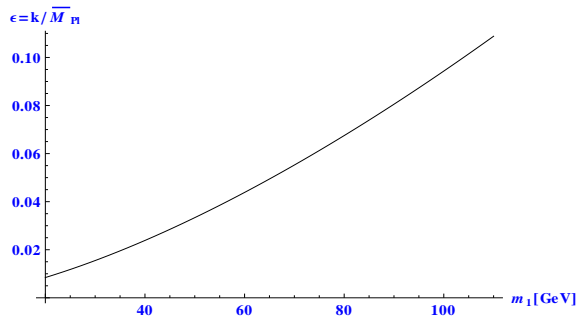}
\caption{figure of $\epsilon$ vs. $m_1$ where $m_H=126\rm{GeV}$ and 
  $m_1$ is varying from 20 GeV to 110 GeV}
\label{atlastheory}
\end{center}
\end{figure}

As argued in \cite{RS}
we assumed that $k<M$ with $M\sim\overline{M}_{Pl}$.
This requirement emerges from the fact that $k$, which measures 
the bulk curvature must be smaller than the Planck scale 
so that the classical solutions for the bulk metric 
given by RS model is a valid one. Alternatively from 
the view point of string theory it was argued in \cite{DRgr} that 
using the expression for D-3 brane tensions and the  string scale 
which in turn is related to $\overline{M}_{Pl}$ through Yang-Mills gauge coupling,
the favoured range of $k/\overline{M}_{Pl}$ is $0.01\leq k/\overline{M}_{Pl}\leq1$.
It has been shown \cite{DRgr}, that in such models
the gravity which propagates in the bulk
can be expanded into KK modes as,\\
\begin{equation}
 h_{\alpha\beta}(x,\phi)=\sum^{\infty}_{n=0}h^{(n)}_{\alpha\beta}(x)
\frac{\chi^{n}(\phi)}{\sqrt r_{c}}\label{grmmex}
\end{equation}
Where $h^{(n)}_{\alpha\beta}(x)$ are the KK modes of the graviton 
on the flat 3-brane and $\chi^{n}(\phi)$ is the wave function for the graviton.
In \cite{DRgr} it has been shown that the solution for the 
graviton wave function is of the form,
\begin{equation}
 \chi^{n}(\phi)=\frac{e^{2\sigma(\phi)}}{N_n}[J_2(z_n)+\alpha_{n}Y_{2}(z_{n})]\label{wfungr}
\end{equation}
Where $J_2$ and $Y_2$ are the Bessel functions of order 2 and
$z_n(\phi)=\frac{m_n}{k}e^{\sigma(\phi)}$, $\alpha_n$ are
constants and $N_n$ is the normalization factor
for the wave function.\\
If we now choose a parameter $x_n=z_n(\pi)$ and consider the 
limit $m_n/k\ll 1$, $e^{kr_c\pi}\gg 1$ then applying the 
continuity condition that the wave function should
be continuous at $\phi=0$ and $\phi=\pi$ (i.e at the two
orbifold fixed points), it produces $J_1(x_n)=0$ and $\alpha_n\ll 1$.
Hence the term with $Y_2(z_n)$ can be ignored in comparison to $J_2(z_n)$.
The expression $J_1(x_n)=0$ implies that
$x_n$ are the roots of the Bessel function of the order of 1 and if we plot the 
$J_1(x_n)$ vs. $x_n$, we get the first few values of $x_n$ as follows,
$x_1=3.83$, $x_2=7.02$, $x_3=10.17$ etc.\\
As it has been defined earlier that, $x_n=z_n(\pi)=\frac{m_n}{k}e^{kr_c\pi}$,
we can find the KK mass tower of graviton from this expression,
\begin{equation}
 m_n=x_nke^{-kr_c\pi}\label{mmgr}
\end{equation}
We mentioned earlier that to study the phenomenology
of this model the most important parameter is $\epsilon=k/\overline{M}_{Pl}$.
Hence let us focus on the first KK mass mode of graviton (i.e $m_1$) and
slightly reparametrize the expression for $m_1$, from eq.(\ref{mmgr}).
\begin{equation}
 m_1=x_1ke^{-kr_c\pi}
\end{equation}
From eq.(\ref{physmass}) and using $\epsilon=k/\overline{M}_{Pl}$,
\begin{equation}
 m_1=x_1\epsilon \frac{m_H}{m_0}\overline{M}_{Pl}\label{1stmmgr}
\end{equation}
Now from eq.(\ref{rplanckmass}), the exponential term 
is very small in comparison to 1. Hence we obtain,
$\overline{M}_{Pl}^2=\frac{M^3}{k}$.
In eq.(\ref{1stmmgr}), we use $\overline{M}_{Pl}^2=\frac{M^3}{k}$
and take the cut off scale as, $m_0=\alpha M$, where 
$M$ is the 5-dimensional Planck scale and $\alpha$ is any constant parameter.
$\alpha=1$ implies that the cut-off scale is the quantum gravity scale
$\overline{M}_{Pl}$, while $\alpha<1$ indicates the appearence of new physics
below Planck scale.
Hence we obtain,
\begin{equation} 
m_1=x_1\epsilon^{2/3}\frac{m_H}{\alpha}\label{1stmmgr2}
\end{equation}
If we consider $\alpha$ to be $1$ i.e 
the cut-off scale is the 5-dimensional Planck scale $M$ then values of 
$m_1$ for different values of $\epsilon$ varying from $0.01-0.1$ 
are shown in table(\ref{t1}) and plotted in figure(\ref{atlastheory}).\\

\begin{figure}[!h]
 \begin{center}
    \centering
    \includegraphics[width=3.40in,height=1.70in]{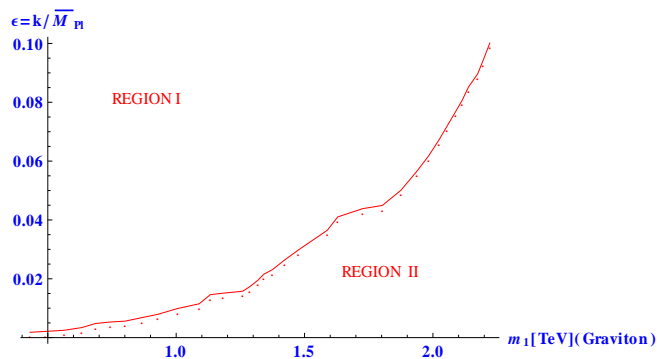}
    \caption{Graph of $\epsilon$ vs. $m_1$ as reported by ATLAS Collaboration.
RegionI has been ruled out which sets lower bound for the first
graviton KK mode for different values of $\epsilon$}
\label{atlasplot}
\end{center}
\end{figure}

\begin{table}[!h]
 \begin{center}
    \centering
\begin{footnotesize}
  \begin{tabular}{||c|c||}\hline
 $\epsilon=\frac{k}{\overline{M}_{Pl}}$  & $m_1=x_1\epsilon^{2/3} m_H$(GeV)\\ \hline
 $0.01$ & $22.39$\\ \hline
 $0.03$ & 46.59\\ \hline
 0.05 & 65.49\\ \hline
 0.07 & 81.96\\ \hline
 0.09 & 96.91\\ \hline
 0.1 & 103.96\\ \hline
  \end{tabular}
\end{footnotesize}
\end{center}
\caption{Theoretical values of first KK mass mode of graviton From RS
 model when $\alpha=1$, $x_1=3.83$ and $m_H=126.0 \rm{GeV}$}
\label{t1} 
\end{table}
The Experimental lower bound for the mass
of the first KK mode of graviton for different values of
$\epsilon$ as reported by the ATLAS Collaboration \cite{ATLAS1} is shown in 
the figure(\ref{atlasplot}). Some of the values of $m_1$
for different $\epsilon$ are shown in table(\ref{t2}).\\ 

\begin{table}[!h]
 \begin{center}
    \centering
\begin{footnotesize}
    \begin{tabular}{||c|c||}\hline
 $\epsilon=k/\overline{M}_{Pl}$  & $m_1$(TeV)\\ \hline
 $0.01$ & $1.01$\\ \hline
 $0.03$ & 1.48\\ \hline
 0.05 & 1.88\\ \hline
 0.07 & 2.04\\ \hline
 0.09 & 2.17\\ \hline
 0.1 & 2.22\\ \hline
 \end{tabular}
\end{footnotesize}
      \caption{The mass table 
from the results of ATLAS
 detector of LHC}
\label{t2}
\end{center}
\end{table}
\subparagraph*{}
The tables as well as the plot clearly indicate that for the entire
range of $0.01<\epsilon<1$, the theoretical prediction for the
first KK mode of graviton falls well within region-I of the 
ATLAS plot which has been ruled out in the search of graviton KK
mode resonance in dilepton events. This implies a serious conflict
between graviton KK modes as predicted in RS model and the result reported by 
ATLAS Collaboration.
\subparagraph*{}
For a possible resolution to this problem we calculate the 
threshold values of the parameter $\alpha$ from the expression,
$\alpha=x_1\epsilon^{2/3}m_H/m_1$ and 
use the values of lower bound of $m_1$ 
for different $\epsilon$ as reported by
ATLAS data. These values are shown in table(\ref{t3}).
\begin{table}[!h]
 \begin{center}
  \begin{tabular}{||c|c|c||}\hline
$\epsilon=k/\overline{M}_{Pl}$ & $m_1$ from ATLAS(TeV) 
& values of $\alpha$ \\ \hline
$0.01$ & 1.01 & $2.2\times 10^{-2}$\\ \hline
0.03 & 1.48 & $3.1\times10^{-2}$\\ \hline
0.05 & 1.88 & $3.4\times10^{-2}$\\ \hline
0.07 & 2.04 & $4.0\times10^{-2}$\\ \hline
0.09 & 2.17 & $4.4\times10^{-2}$\\ \hline
0.1 & 2.22 & $4.6\times 10^{-2}$\\ \hline
\end{tabular}
 \end{center}
\caption{The values of $\alpha$ for each $\epsilon$ and corresponding mass
of graviton from ATLAS data where $x_1=3.83$ and $m_H=126$ GeV}
\label{t3}
\end{table}\\
This indicates that the cut-off of the RS model namely $m_0$,
must be at least two order lower then the Planck scale 
indicating the existence of new physics at a scale of the order of 
$10^{17}$ GeV.
\section*{Conclusion}
Present estimation of the lower bound on the lightest RS graviton KK mode masses
from dilepton events as reported by ATLAS Collaboration
is in conflict with the requirement of resolving the gauge hierarchy
/fine tuning problem unless the cut-off of the 
standard model is atleast two order lower than the Planck/Quantum gravity scale.
This indicates the possible appearence of a new physics at
$\sim 10^{17}$ GeV.


\end{document}